\documentclass{PoS}

\title{   
\begin{picture}(0,0)(0,0)%
   \put(230,75){\makebox(0,0)[l]{\textnormal
{\normalsize OU-HET-732-2011, UTHEP-636, KEK-CP-259
}}}%
 \end{picture}%
Chiral interpolation in a finite volume}

\ShortTitle{Chiral interpolation}

\author{JLQCD Collaboration: 
        \speaker{H.~Fukaya}$^a$\thanks{E-mail: hfukaya@het.phys.sci.osaka-u.ac.jp},
        S.~Aoki$^{b,c}$,
        S.~Hashimoto$^{d,e}$,
        T.~Kaneko$^{d,e}$,
        H.~Matsufuru$^{d,e}$,
        J.~Noaki$^{d}$,
        T.~Onogi$^a$
        and
        N.~Yamada$^{d,e}$
        \\
        \\
        \\
        \llap{$^a$}
        Department of Physics, Osaka University, 
        Toyonaka, Osaka 560-0043 Japan
        \\
        \llap{$^b$}
        Graduate School of Pure and Applied Sciences, 
        University of Tsukuba, Tsukuba 305-8571, Japan
        \\ 
        \llap{$^c$}
        Center for Computational Sciences, University of Tsukuba, 
        Tsukuba 305-8577, Japan
        \\
        \llap{$^d$}
        High Energy Accelerator Research Organization (KEK),
        Tsukuba 305-0801, Japan 
        \\
        \llap{$^e$}
        School of High Energy Accelerator Science,
        The Graduate University for Advanced Studies (Sokendai),
        Tsukuba 305-0801, Japan
}

\abstract{
A simulation of lattice QCD at (or even below) 
the physical pion mass is feasible 
on a small lattice size of $\sim$ 2 fm. 
The results are, however, subject to large finite volume effects. 
In order to precisely understand the chiral behavior in a finite volume, 
we develop a new computational scheme  
to interpolate the conventional $\epsilon$ and $p$ regimes 
within chiral perturbation theory. 
In this new scheme, we calculate the two-point function in the pseudoscalar channel, 
which is described by a set of Bessel functions in an infra-red 
finite way as in the $\epsilon$ regime, while chiral logarithmic effects 
are kept manifest as in the $p$ regime. The new ChPT formula is compared 
to our 2+1-flavor lattice QCD data near the physical up and down quark mass, 
$m_{ud}\sim$ 3 MeV on an $L\sim 1.8$ fm lattice.
We extract the pion mass $= 99(4)$ MeV, 
from which we attempt a chiral ``interpolation'' 
of the observables to the physical point.
}

\FullConference{ The XXIX International Symposium on Lattice Field Theory - Lattice 2011\\
July 10-16, 2011\\
Squaw Valley, Lake Tahoe, California}

\begin{document}

\section{Introduction}


In the standard lattice QCD studies, one tries to
keep the volume size $L$ (or $V^{1/4}$) large 
so that the physics does not change very much from
its infinite volume limit.
Namely, denoting the generic pion mass by $M$,
a dimensionless combination $ML$ must be large.
However, the numerical cost sharply grows
in such a scaling limit, $M\to 0$ keeping a large value of $ML$.

In this work, we propose an alternative way:
to investigate the chiral limit in a fixed sized box.
With a fixed value of the lattice size,
the numerical cost scales much mildly with $M$
and even it saturates in the vicinity of the chiral limit
since the lowest eigenvalue of the Dirac operator
is no more controlled by $1/M$ but has a gap controlled by $1/V$.
In fact, the JLQCD and TWQCD collaborations have been performing
lattice QCD simulations near the chiral limit
with dynamical overlap quarks 
\cite{Fukaya:2007fb, Fukaya:2009fh, Fukaya:2010na}.

The results are, of course, largely 
distorted from those in the infinite volume limit.
It is, however, possible to analytically correct
the finite size scaling
using chiral perturbation theory (ChPT)
\cite{Gasser:1983yg, Gasser:1984gg}
since only the pion has a long correlation length
near the chiral limit.
If one has a good control of the pion physics,
one can convert the data on a finite size lattice
to those in the large volume limit,
as long as the size of the system
$L$ is well above the inverse QCD scale $1/\Lambda_{\rm QCD}$.

In the very vicinity of the chiral limit,
the $\epsilon$ expansion of ChPT \cite{Gasser:1986vb}
is useful
as it treats the zero-momentum mode non-perturbatively, 
which gives the dominant contribution to the finite size effects.
But the $\epsilon$ expansion is valid only in a small
range $ML \ll 1$ (called the $\epsilon$ regime)
and the formulas look very different
from those in the  conventional $p$ regime.
As we want to analyze the data both in and out of the 
$\epsilon$ regime in a uniform way, the use of 
the $\epsilon$ expansion is not very suitable.

Recently, a new perturbative approach of chiral expansion 
which 
interpolates the $p$ and $\epsilon$ expansions is 
proposed \cite{Damgaard:2008zs}
and the calculation is extended to the two-point functions by
two of the authors \cite{Aoki:2011pz}.
This new scheme has no limitation on $ML$.
The calculation is done by keeping both features of 
$p$ and $\epsilon$ expansions: treating the zero-mode separately
and exactly, while keeping all the terms that appear in the
$p$ expansion\footnote{Note that, at a given order of expansion, 
the $\epsilon$ expansion has less terms 
than the $p$ expansion because
it treats the mass term as one order smaller perturbation.}.
The results are expressed by a set of Bessel functions 
in an infra-red finite way as in the $\epsilon$ expansion,
while the chiral logarithmic effects are kept manifest
as in the $p$ expansion.

Here we review the new perturbative scheme of ChPT
and compare the formula with (preliminary) lattice QCD data.
Our data at the lightest up and down quark mass $m_{ud}\sim 3$ MeV,
indicate the pion mass less than its physical value.
This means that in our finite size lattice,
one can attempt a chiral ``interpolation''
of the observables to the physical point. 
 
\section{New chiral expansion}

The difference between the two conventional $p$ and $\epsilon$ expansions
is in their parametrization of the chiral field 
(we denote $U(x)\in SU(N)$)
and the counting rule for the mass term.
In the $p$ expansion, both of the zero mode and 
the non-zero momentum modes are equally and perturbatively treated,
and the mass term appears as a leading-order (LO) term in the Lagrangian.
Namely, 
\begin{eqnarray}
\mbox{$p$ expansion} &:& U(x)=\exp\left(i\frac{\sqrt{2}\xi(x)}{F}\right),
\;\;\;\;\;
M\sim {\cal O}(1/L),
\end{eqnarray}
where $\xi(x)$ denotes the generic pion field, and $F$ is the (bare)
pion decay constant. The counting rule is given in the units of 
the smallest non-zero momentum $1/L$.

In the $\epsilon$ regime, we treat the zero-mode separately
and non-perturbatively, while the
mass term is treated as a next-to-leading order (NLO) correction.
Namely, we have 
\begin{eqnarray}
\mbox{$\epsilon$ expansion} &:& 
U(x)=U_0\exp\left(i\frac{\sqrt{2}\bar{\xi}(x)}{F}\right),
\;\;\;\;\;
M\sim {\cal O}(1/L^2),
\end{eqnarray}
where $U_0$ denotes the zero-mode for which 
an exact group integration is performed over $SU(N)$ 
(or $U(N)$ in a fixed topological sector) manifold.
Note that the zero-mode is absent in the perturbative mode 
$\bar{\xi}(x)$ (a condition $\int d^4 x\;\bar{\xi}(x)=0$ is imposed).

For our new computational scheme, which let us 
denote ``$i$'' (=interpolating) expansion, 
we have to keep the both features, non-perturbative treatment
of the zero momentum mode, and the mass term kept
at LO:
\begin{eqnarray}
\mbox{$i$ expansion} &:& 
U(x)=U_0\exp\left(i\frac{\sqrt{2}\bar{\xi}(x)}{F}\right),
\;\;\;\;\;
M\sim {\cal O}(1/L).
\end{eqnarray} 

In Refs.\cite{Damgaard:2008zs, Aoki:2011pz}, 
an additional counting rule is given for
a certain combination of the quark mass matrix and 
the zero-mode : $\mathcal{M}(U_0-1)\sim {\cal O}(1/L^3)$,
which helps to identify relevant/irrelevant diagrams.
One can justify this new rule by a direct group integration 
(See Refs.\cite{Damgaard:2008zs, Aoki:2011pz} for the details.).

With this new perturbative scheme, one can calculate
correlation functions.
As shown in Ref.~\cite{Aoki:2011pz},
the calculation, which is a mixture of matrix integrals
and perturbative $\xi$ integrals,
 is fairly tedious but straightforward.
In the end, one obtains a simple form for 
the calculation for the pseudoscalar correlator, 
\begin{eqnarray}
\label{eq:PP}
\int d^3 x\langle P({\bf x},t)P(0,0)\rangle 
&=& A\frac{\cosh (m_\pi^{\mbox{\scriptsize 1-loop}}(t-T/2))}{\sinh(m_\pi^{\mbox{\scriptsize 1-loop}}T/2)}+B,
\end{eqnarray}
where $T$ denotes the temporal extent of the volume,
and $m_\pi^{\mbox{\scriptsize 1-loop}}$ denotes the pion mass 
which contains one-loop corrections including
the finite size effects (from the non-zero modes),
as well as $Q$ dependence if the topology is fixed.

The coefficient $A$ is a function of
$m_\pi^{\mbox{\scriptsize 1-loop}}$, the decay constant
 $f_\pi^{\mbox{\scriptsize 1-loop}}$, and the 
chiral condensate $\Sigma_{\rm eff}$ while
the constant $B$ is a function of $\Sigma_{\rm eff}$ only (through a dimensionless combination $\mathcal{M}\Sigma_{\rm eff} V$).
These physical parameters include one-loop corrections from non-zero modes
and thus the chiral logarithms as in the $p$ regime.
$\Sigma_{\rm eff}$ dependence is, however, embedded 
through the modified Bessel functions in an infra-red finite way,
just as in the $\epsilon$ expansion.

In this new formula (\ref{eq:PP}), the constant
term $B$ plays an essential role in the interpolation 
between the $\epsilon$ and $p$ regimes: 
it precisely cancels an unphysical infra-red divergence 
in the first term in the massless limit,
while it rapidly disappears in the large mass region.
One can confirm that the $\epsilon$ expansion formula 
and that 
in the $p$ expansion 
are interpolated
as a smooth function of $m_\pi^{\mbox{\scriptsize 1-loop}}$
by the new formula.

\section{Preliminary lattice results}

Let us compare the new ChPT formula with 
the lattice data generated by
the JLQCD and TWQCD collaborations.
Numerical simulations are performed with
the Iwasaki gauge action at $\beta=2.3$ including 2+1 flavors of
dynamical overlap quarks on a $16^3\times 48$ lattice.
The lattice cutoff $1/a$ = 1.759(8)(5)~GeV is determined from
the $\Omega$-baryon mass.

For the strange quark mass, we choose two different values 
but here we concentrate on the data at $m_s$ = 0.080,
which is closer to the physical value $0.081$ determined from the kaon mass.
With this fixed value of $m_s$, five values of up and down quark mass
$m_{ud}$ = 0.002, 0.015, 0.025, 0.035, and 0.050 are taken.
The smallest value $m_{ud}=0.002$
roughly corresponds to 3~MeV in the physical unit 
($\overline{\rm MS}$ at $2$ GeV), 
where the pions are in the $\epsilon$ regime 
while the kaons still remain in the $p$ regime. 

In the Hybrid Monte Carlo (HMC) updates, the global topological
charge $Q$ of the gauge field is fixed to its initial value by
introducing extra (unphysical) Wilson fermions, 
which have a negative mass of cutoff order.
In our main runs presented here, we set $Q=0$.

For the computation of the pseudoscalar correlator,
we use smeared sources with the form of a single exponential function.
We observe that the smearing is effective even in the 
$\epsilon$ regime.
To improve the statistical signal, the low-mode averaging
technique is used: the low-mode part of the correlator 
is separately calculated by the 80 eigenmodes of the Dirac operator 
and averaged over different source points.

The auto-correlation time of the simulation is estimated by
the history of the lowest Dirac eigenvalue, 
which turns out to be 6--24 trajectories depending 
on the simulation parameters.
The statistical error is estimated by the jackknife method
after binning data in every 100 trajectories.

Details of the numerical simulation will be reported elsewhere.

We attempt a two parameter ($m_\pi^{\mbox{\scriptsize 1-loop}}$ 
and $f_\pi^{\mbox{\scriptsize 1-loop}}$) fit with the fit function 
\begin{eqnarray}
\label{eq:fit}
f(t;\; m_\pi^{\mbox{\scriptsize 1-loop}}, f_\pi^{\mbox{\scriptsize 1-loop}}) 
&=& A(m_\pi^{\mbox{\scriptsize 1-loop}},f_\pi^{\mbox{\scriptsize 1-loop}},
\Sigma_{\rm eff})
\frac{\cosh(m_\pi^{\mbox{\scriptsize 1-loop}}(t-T/2))}
{\sinh (m_\pi^{\mbox{\scriptsize 1-loop}} T/2)}
+B(\Sigma_{\rm eff}),
\end{eqnarray}
taking
$\Sigma_{\rm eff}=0.00204(07)$ from our recent result \cite{Fukaya:2010na},
as the input.

In order to determine the fitting range of $t$,
we define the ``local'' mass and decay constant
$m_\pi^{lc}(t)$ and $f_\pi^{lc}(t)$ at each time slice $t$,
by the solutions of the equations
\begin{eqnarray}
f(t;\; m^{lc}_\pi(t), f^{lc}_\pi(t))&=&\mbox{lattice data at}\;\;\;t,\\
f(t+1;\; m^{lc}_\pi(t), f^{lc}_\pi(t))&=&\mbox{lattice data at}\;\;\;t+1,
\end{eqnarray}
which can be numerically solved. 
The two equations are non-linear but we confirm
that the solution for a given $t$ is unique, 
at least in a range 
$m^{lc}_\pi(t),  f_\pi^{lc}(t) < \mbox{2GeV}$.

\begin{figure*}
\centering
\includegraphics[width=7.5cm]{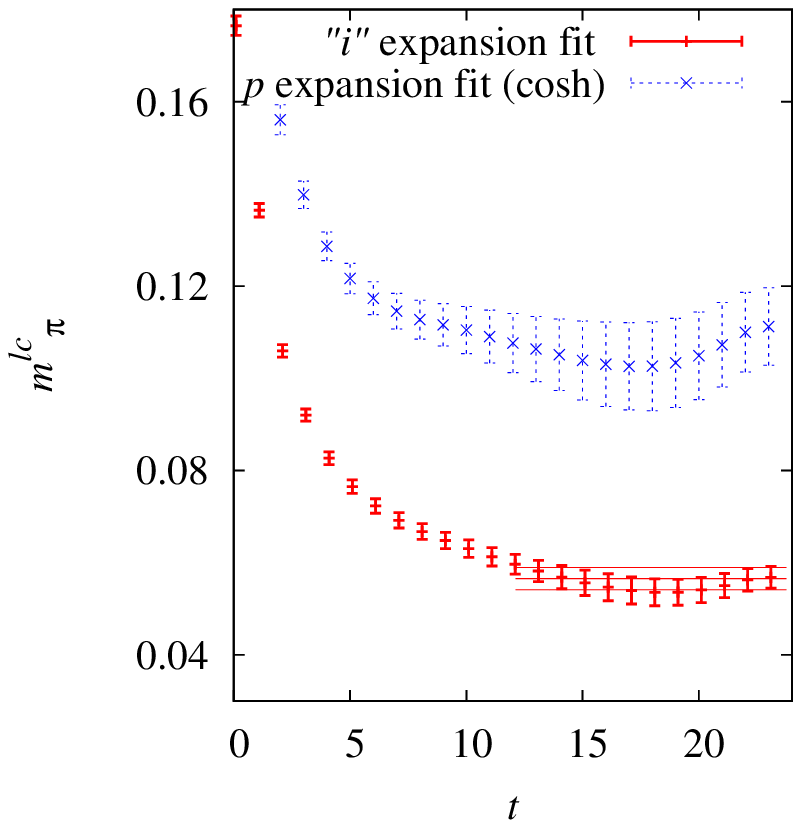}
\includegraphics[width=7.5cm]{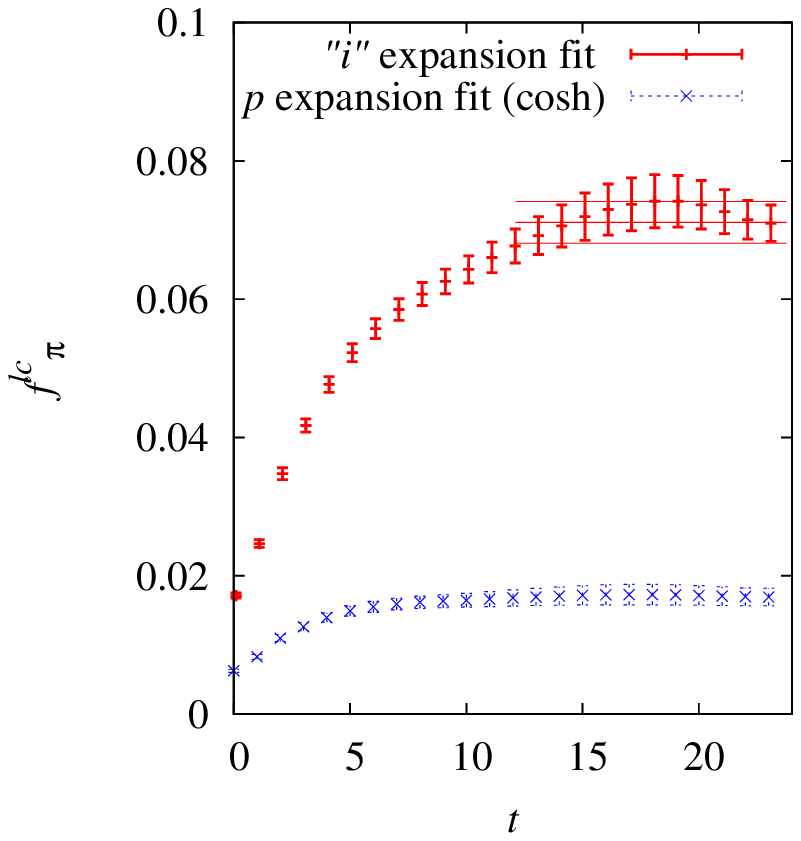}
\caption{The ``local'' mass (left) and decay constant (right) of the pion
at each time slice. 
The cross symbols show the conventional effective mass using the simple $\cosh$ form.}
\label{fig:effMF}
\end{figure*}

Figure~\ref{fig:effMF} shows the local mass and decay constant plots
for the lightest mass ($m_{ud}=0.002$) data.
We find a plateau 
for both of the mass and decay constant, from which
we determine the fitting range as $t\geq 13$.
For comparison, we also present a plot obtained 
with the conventional $p$ expansion formula, which yields an
unreasonable value $f^{lc}_\pi\sim 0.017 (30 \mbox{MeV})$.
The zero-mode effect is thus shown to be essential 
near the chiral limit.

From the fit in the range $t\geq 13$,
we obtain $(m_\pi^{\mbox{\scriptsize 1-loop}}, f_\pi^{\mbox{\scriptsize 1-loop}}) 
$$= (0.0566(24), 0.0711(30))$.
We can further use ChPT to correct
perturbative finite volume effects 
from the non-zero momentum modes.
After correcting these finite size effects with ChPT at one-loop,
we obtain 
\begin{eqnarray}
m_\pi^{\mbox{\scriptsize 1-loop}}|_{V\to\infty} 
&=& 0.0561(24)\;[98.7(4.2)\;\mbox{MeV}],\\
f_\pi^{\mbox{\scriptsize 1-loop}}|_{V\to\infty} 
&=& 0.0724(30)\;[127.0(5.3)\;\mbox{MeV}],
\end{eqnarray}
at $m_{ud}=0.002$.
This result implies that 
the quark mass for our simulation in the $\epsilon$ regime
is below the physical point.

We can repeat the same analysis also for the larger mass
data in the $p$ regime.
However, we find that the results are different by only
$1$\% from our previous $p$ expansion analysis \cite{Noaki:2010zz}.
The zero-momentum mode effects in this region, $m_{ud}\geq 0.015$,
are thus negligible.

Since $m_\pi^{\mbox{\scriptsize 1-loop}}|_{V\to\infty}$ and 
$f_\pi^{\mbox{\scriptsize 1-loop}}|_{V\to\infty}$ should have
the conventional logarithmic $m_{ud}$ dependence,
we can ``interpolate'' the data to the physical value.
Here we attempt the following two methods:
\begin{enumerate}
\item[1.] SU(3) ChPT 3-point ($m_{ud}=0.002,0.015,0.025$) combined ($m_\pi$ and $f_\pi$ simultaneously) fit
with 4 free parameters (the chiral condensate $\Sigma_0$ 
and decay constant $f_0$ both in the
$m_{ud}=0$ limit, and 
the NLO low-energy constants
$L^r_M\equiv 2L_6^r+L_8^r$, and $L_F^r \equiv 2L_4^r+L_5^r$).
\item[2.] linear 3-point fit for each of $m_\pi$ and $f_\pi$.
\end{enumerate}

\begin{figure}
\centering
\includegraphics[width=7.5cm]{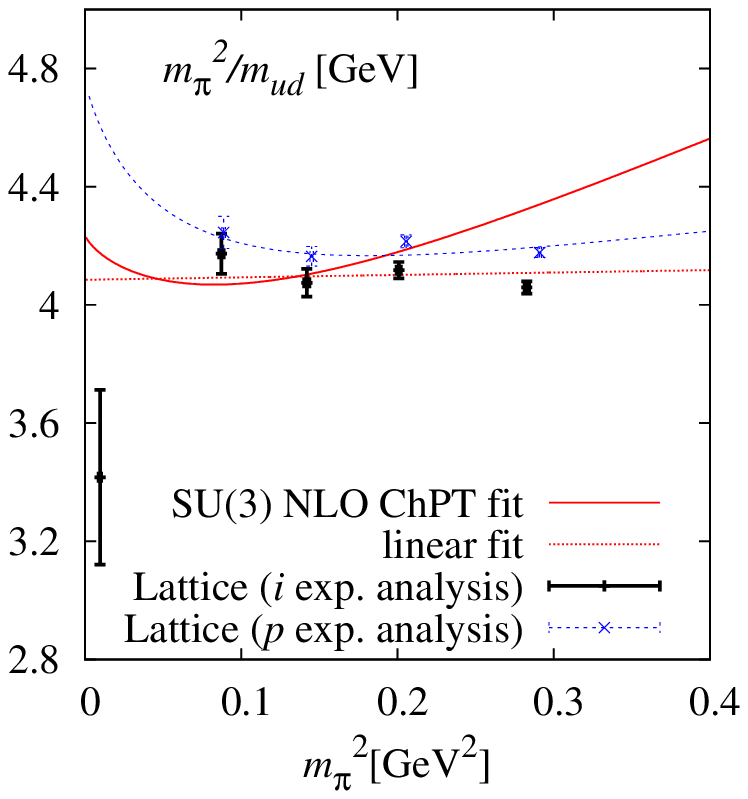}
\includegraphics[width=7.5cm]{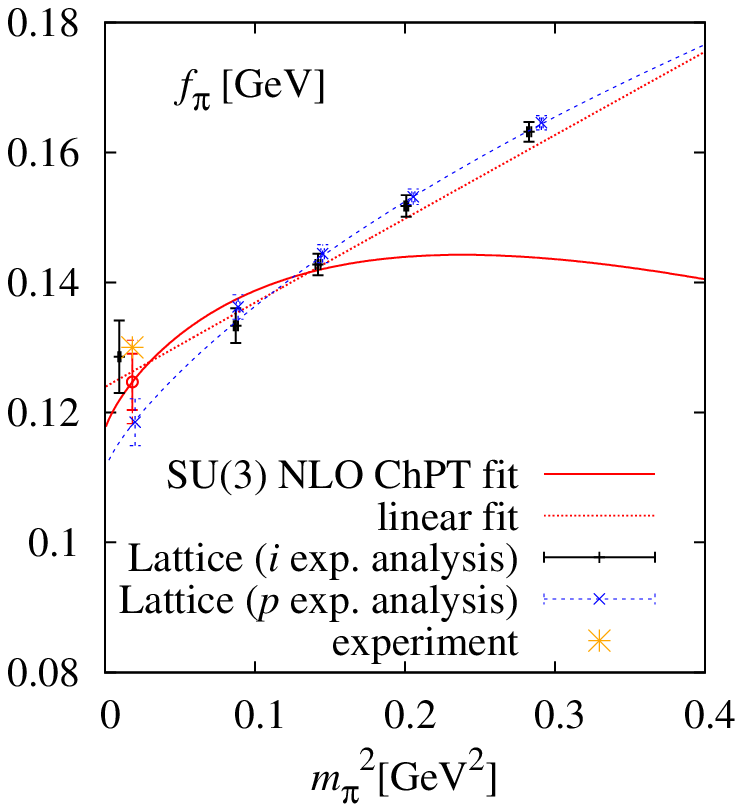}
\caption{The chiral interpolation of the pion mass (left) 
and decay constant (right).
}
\label{fig:CI}
\end{figure}

The results are shown in Fig.~\ref{fig:CI}.
The physical point of $f_\pi$ (open circle) is 
determined from the experimental input 
$m_\pi=135$ MeV.
Although the interpolated value is consistent 
with the experimental value,
our data do not show the striking effect of the chiral logarithm.
The lightest pion mass looks too low, 
by $\sim 7.5$\%
compared to the expected ChPT curve.
The linear fit for the pion decay constant 
looks better. Similar result was also reported by
RBC-UKQCD collaboration \cite{Aoki:2010dy}.
For comparison, our previous study in the $p$ regime
is also presented in Fig.~\ref{fig:CI} (dashed curves).

From the 4 parameter SU(3) fit,
we extract the pion decay constant (at $m_s=0.08$) as
\begin{eqnarray}
f_\pi (\mbox{ at physical}\;m_{ud}) &=& 125(4)(^{+5}_{-0}) \mbox{ MeV},\\
f_0 (m_{ud}=0) &=& 117(5)(^{+8}_{-0}) \mbox{ MeV},
\end{eqnarray}
where the first error is statistical and the second is
systematic error from the choice of the interpolation function.

\section{Convergence of the new formula}

Our new ChPT formula is expected to be valid up to NNLO corrections.
In order to estimate the systematic effects
from the higher order contribution, we reanalyze
our lattice data with another expression of the formula:
\begin{eqnarray}
\label{eq:PPmod}
\int d^3 x\langle P({\bf x},t)P(0,0)\rangle 
&=& \left[\frac{A}{\sinh(m_\pi^{\mbox{\scriptsize 1-loop}}T/2)}+B\right]\cosh (Z_m m_\pi^{\mbox{\scriptsize 1-loop}}(t-T/2)),\nonumber\\
Z_m&=&\sqrt{\frac{1}{1+B\sinh(m_\pi^{\mbox{\scriptsize 1-loop}}T/2)/A}},
\end{eqnarray}
where $A$ and $B$ are the same constants 
as those in (\ref{eq:PP}).
Noting that $B$ rapidly disappears in the $p$ regime,
one can confirm that the formula (\ref{eq:PPmod}) 
is equivalent to (\ref{eq:PP}) up to NNLO corrections.

For the $p$ regime data $m_{ud}\geq 0.015$, 
the change of the formula gives only 1\% level differences.
Namely, the formulas (\ref{eq:PP}) and (\ref{eq:PPmod})
are equally good and NNLO contribution is well under control.

However, for the lightest mass case $m_{ud}=0.002$, 
we obtain $m_\pi^{\mbox{\scriptsize 1-loop}}=0.0636(32)$,
which is 12\% higher than the original analysis (0.0566(24)),
while $f_\pi^{\mbox{\scriptsize 1-loop}}=0.0732(41)$ is consistent within 
the statistical error.

The significant difference of $m_\pi$ 
in the $\epsilon$-regime data only,
may be explained as follows.
Since we separately treat the zero mode and non-zero modes,
our ChPT expansion for a general quantity has a form
\begin{eqnarray}
O &=& O_{\rm LO+NLO} +  \delta O_{\rm NNLO}(m_\pi^2, 1/V), 
\end{eqnarray}
where the higher order correction term $\delta O_{\rm NNLO}(m_\pi^2, 1/V)$
has ``mass-independent'' contributions.
For $m_\pi$, the LO+NLO contribution decreases 
for the lower quark mass while the correction term is kept finite,
which means that the formula is less sensitive to $m_\pi$ 
in the low mass region.
This is not surprising since $m_\pi$ is treated as an NLO quantity 
in the $\epsilon$ expansion. 
On the other hand, the determination of the 
decay constant $f_\pi$ is stable 
as it has a finite chiral limit and treated as LO
even in the $\epsilon$ expansion.

\section{Summary}

We have developed a new computational scheme in ChPT
which interpolates the conventional $\epsilon$ and $p$ regimes.
The new formula for the pseudoscalar correlator
 allows us to analyze the lattice data
both in the $\epsilon$ and $p$ regimes equally and simultaneously.

Simulating the physical quark mass on the lattice is feasible within a
reasonable computational cost if the volume is kept small.
If we have a good control of the pion zero-mode within ChPT,
we can precisely estimate the physical values in the large volume limit,
and attempt a chiral interpolation to the physical point.
In this work, we have demonstrated that the chiral interpolation indeed works
for determination of the pion decay constant,
while the pion mass seems to have a bad sensitivity to our scheme. 

We thank P.~H.~Damgaard and C.~Lehner for useful discussions.
Numerical simulations are performed on the IBM System Blue Gene
Solution at High Energy Accelerator Research Organization
(KEK) under a support of its Large Scale Simulation
Program (No. 09/10-09). 
This work is supported in part by the Grant-in-Aid of the
  Japanese Ministry of Education
  (No.21674002, 21684013, 23105710), 
the Grant-in-Aid for Scientific
Research on Innovative Areas (No. 2004: 20105001, 20105002, 20105003, 20105005),
and the HPCI Strategic Program of Ministry of Education.

\end{document}